\documentclass[epj]{webofc}
\usepackage[varg]{txfonts}   
\wocname{EPJ Web of Conferences}
\woctitle{INPC 2013}
\begin{document}
\title{Early Results from the Q$_{\rm weak}$ Experiment}

\author{
D.~Androic\inst{1}
\and 
D.S.~Armstrong\inst{2}
\and 
A.~Asaturyan\inst{3}
\and
T.~Averett\inst{2}
\and
J.~Balewski\inst{4}
\and
J.~Beaufait\inst{5}
\and
R.S.~Beminiwattha\inst{6}
\and
J.~Benesch\inst{5}
\and
F.~Benmokhtar\inst{7}
\and
J.~Birchall\inst{8}
\and
R.D.~Carlini\inst{5}
\and
G.D.~Cates\inst{9}
\and
J.C.~Cornejo\inst{2}
%
\and
S.~Covrig\inst{5} \and
M.M.~Dalton\inst{9} \and
C.A.~Davis\inst{10} \and
W.~Deconinck\inst{2} \and
J.~Diefenbach\inst{11} \and
J.F.~Dowd\inst{2} \and
J.A.~Dunne\inst{12} \and
D.~Dutta\inst{12} \and
W.S.~Duvall\inst{13} \and
M.~Elaasar\inst{14} \and
W.R.~Falk\inst{8} \and
J.M.~Finn\inst{2} \and
T.~Forest\inst{15,16}  \and
D.~Gaskell\inst{5} \and
M.T.W.~Gericke\inst{8} \and
J.~Grames\inst{5} \and
V.M.~Gray\inst{2} \and
K.~Grimm\inst{16,2} \and
F.~Guo\inst{4} \and
J.R.~Hoskins\inst{2} \and
K.~Johnston\inst{16} \and
D.~Jones\inst{9} \and
M.~Jones\inst{5} \and
R.~Jones\inst{17} \and
M.~Kargiantoulakis\inst{9} \and
P.M.~King\inst{6} \and
E.~Korkmaz\inst{18} \and
S.~Kowalski\inst{4} \and
J.~Leacock\inst{13} \and
J.~Leckey\inst{2} \fnsep\thanks{now at Indiana University, Bloomington, Indiana 47405, USA} \and
A.R.~Lee\inst{13} \and
J.H.~Lee\inst{6,2} \fnsep\thanks{now at Institute for Basic Science, Daejeon, South Korea} \and
L.~Lee\inst{10,8} \and
S.~MacEwan\inst{8} \and
D.~Mack\inst{5} \and
J.A.~Magee\inst{2} \and
R.~Mahurin\inst{8} \and
J.~Mammei\inst{13} \and
J.~Martin\inst{19} \and
M.J.~McHugh\inst{20} \and
D.~Meekins\inst{5} \and
J.~Mei\inst{5} \and
R.~Michaels\inst{5} \and
A.~Micherdzinska\inst{20} \and
A.~Mkrtchyan\inst{3} \and
H.~Mkrtchyan\inst{3} \and
N.~Morgan\inst{13} \and
K.E.~Myers\inst{20} \fnsep\thanks{now at Rutgers, the State University of New Jersey, Piscataway, NJ 08854 USA} \and
A.~Narayan\inst{12} \and
L.Z.~Ndukum\inst{12} \and
V.~Nelyubin\inst{9} \and
Nuruzzaman\inst{11,12} \and
W.T.H van Oers\inst{10,8} \and
A.K.~Opper\inst{20} \and
S.A.~Page\inst{8} \and
J.~Pan\inst{8} \and
K.~Paschke\inst{9} \and
S.K.~Phillips\inst{21} \and
M.L.~Pitt\inst{13} \and
M.~Poelker\inst{5} \and
J.F.~Rajotte\inst{4} \and
W.D.~Ramsay\inst{10,8} \and
J.~Roche\inst{6} \and
B.~Sawatzky\inst{5} \and
T.~Seva\inst{22} \and
M.H.~Shabestari\inst{12} \and
R.~Silwal\inst{9} \and
N.~Simicevic\inst{16} \and
G.R.~Smith\inst{5} \fnsep\thanks{\email{smithg@jlab.org}} \and
P.~Solvignon\inst{5} \and
D.T.~Spayde\inst{23} \and
A.~Subedi\inst{12} \and
R.~Subedi\inst{20} \and
R.~Suleiman\inst{5} \and
V.~Tadevosyan\inst{3} \and
W.A.~Tobias\inst{9} \and
V.~Tvaskis\inst{19} \fnsep\thanks{now at University of Manitoba, Winnipeg, MB R3T2N2 Canada} \and
B.~Waidyawansa\inst{6} \and
P.~Wang\inst{8} \and
S.P.~Wells\inst{16} \and
S.A.~Wood\inst{5} \and
S.~Yang\inst{2} \and
R.D.~Young\inst{24} \and
S.~Zhamkochyan\inst{3} 
%
}

\institute{ Department of Physics, University of Zagreb, Zagreb, HR 10002 Croatia 
\and College of William and Mary, Williamsburg, VA 23185 USA
\and A.~I.~Alikhanyan National Science Laboratory (Yerevan Physics and), Yerevan 0036, Armenia
\and Massachusetts and of Technology,  Boston, MA 02139 USA
\and Thomas Jefferson National Accelerator Facility, Newport News, VA 23606 USA
\and Ohio University, Athens, OH 45701 USA
\and Christopher Newport University, Newport News, VA 23606 USA
\and University of Manitoba, Winnipeg, MB R3T2N2 Canada
\and University of Virginia,  Charlottesville, VA 22908 USA
\and TRIUMF, Vancouver, BC V6T2A3 Canada
\and Hampton University, Hampton, VA 23668 USA
\and Mississippi State University,  Mississippi State, MS 39762  USA
\and Virginia Polytechnic and \& State University, Blacksburg, VA 24061 USA
\and Southern University at New Orleans, New Orleans, LA 70126 USA
\and Idaho State University, Pocatello, ID 83209 USA
\and Louisiana Tech University, Ruston, LA 71272 USA
\and University of Connecticut,  Storrs-Mansfield, CT 06269 USA
\and University of Northern British Columbia, Prince George, BC V2N4Z9 Canada
\and University of Winnipeg, Winnipeg, MB R3B2E9 Canada
\and George Washington University, Washington, DC 20052 USA
\and University of New Hampshire, Durham, NH 03824 USA
\and Hendrix College, Conway, AR 72032 USA
\and University of Adelaide,  Adelaide, SA 5005 Australia
}

\abstract{
A subset of results from the recently completed Jefferson Lab Q$_{\rm weak}$ experiment are reported. This experiment, sensitive to  physics beyond the Standard Model, exploits the small parity-violating asymmetry in elastic $\vec{e}$p scattering to provide the first determination of the proton’s weak charge $Q_w^p$. The experiment employed a 180 $\mu$A longitudinally polarized 1.16 GeV  electron beam on a 35 cm long liquid hydrogen target. Scattered electrons in the angular range $6^\circ < \theta  < 12^\circ$ corresponding to Q$^2$ = 0.025 GeV$^2$ were detected in eight Cerenkov detectors arrayed symmetrically around the beam axis. The goals of the experiment were to provide a measure of $Q_w^p$  to 4.2\% (combined statistical and systematic error), which implies a measure of sin$^2$($\theta_w$) at the level of 0.3\%, and to help constrain the vector weak quark charges C$_{1u}$ and C$_{1d}$. The experimental method is described, with particular focus on the challenges associated with the world’s highest power LH$_2$ target. The new constraints on  C$_{1u}$ and C$_{1d}$ provided by the subset of the experiment’s data analyzed to date will also be shown, together with the extracted weak charge of the neutron. 
}

\maketitle

\section{Introduction}
\label{intro}
We report the results obtained from the analysis of data collected during the commissioning run of the Q$_{\rm weak}$ experiment~\cite{Qweak}  performed at Jefferson Lab (JLab). The experiment provides a precise measure of the $\vec{\rm e}$p scattering asymmetry at low Q$^2$. While representing only about 4\% of the total data collected in the experiment, the commissioning data presented here already provide the most precise  $\vec{\rm e}$p 
parity-violating electron scattering (PVES) asymmetry ever measured.  Combined with the small Q$^2$ chosen for the experiment, and including the results of less precise, higher Q$^2$ data to constrain hadronic corrections, a reliable extraction of the threshold quantity $Q^{p}_{W}$ is obtained for the first time.  The weak charge of the proton ($Q^{p}_{W}$) is the neutral-weak analog of the proton's electric charge~\cite{SMQweak}. 

For a target with Z protons and N neutrons, the weak charge can be expressed in terms of the axial electron, vector quark weak charges of the up and down quarks $C_{1i} = 2 g^e_A g^i_V$ according to 
$Q_{w}(Z,N) = -2(C_{1u}(2Z+N) + C_{1d}(Z+2N))$~\cite{PDG2012}. For the proton target used in the experiment reported here, $Q_{w}(p) = -2(2C_{1u} + C_{1d})$. However, in order to extract the two unknown vector quark weak charges $C_{1u}$ and $C_{1d}$, a second equation is required. Precise measurements of atomic parity violation (APV)  in $^{133}$Cs~\cite{APV1} provide this second equation: $Q_{w}(^{133}Cs) = -2(188C_{1u} + 211C_{1d})$. Finally, the resulting vector quark weak charges can in turn be used to determine the weak charge of the neutron: $Q_{w}(n) = -2(C_{1u} + 2C_{1d})$.

\section{Formalism}
\label{formalism}
The asymmetry measured in the experiment is the difference over the sum of the elastic $\vec{\rm e}$p scattering cross section for electrons with positive and negative helicity, 
\begin{equation}
 A_{ep} = {{\sigma_{+}-\sigma_{-}} \over {\sigma_{+}+\sigma_{-}}} .
 \label{arc}   
\end{equation}
This asymmetry may be described at tree level in terms of electromagnetic, weak, and axial form factors as 
\begin{equation}
A_{ep} = A_0
\left[
\varepsilon G^{\gamma}_{\scriptscriptstyle{E}} G^{Z}_{\scriptscriptstyle{E}} 
+ \tau G^{\gamma}_{\scriptscriptstyle{M}}G^{Z}_{\scriptscriptstyle{M}} 
- (1-4 \sin^2 \theta_W ) \varepsilon^{\prime} G^{\gamma}_{\scriptscriptstyle{M}} G^{Z}_{\scriptscriptstyle{A}}
\over 
\varepsilon (G^{\gamma}_{\scriptscriptstyle{E}})^{2} + \tau (G^{\gamma}_{\scriptscriptstyle{M}})^{2}
\right]
 \label{arc2}     
\end{equation}
where
\begin{equation}
A_0 = \frac{- G_F Q^2 }{ 4 \pi \alpha \sqrt{2}}, \; \;
\varepsilon = {1 \over 1 + 2(1 + \tau)\tan^2{\theta \over 2}}, \; \; \rm{and} \; \;
\varepsilon^{\prime} = \sqrt{\tau (1+\tau) (1- \varepsilon^2)}
\label{arc3}    
\end{equation}

\noindent are kinematic quantities, $G_F$ the Fermi constant, $\sin^2 \theta_W$ the weak mixing angle,
 $-Q^2$  the four-momentum transfer squared, $\alpha$ the fine structure constant, $\tau = Q^{2}/4M^{2}$,
$M$ the proton mass,  and $\theta$ the laboratory 
electron scattering angle. 

It's convenient~\cite{2007C1paper} to rewrite Eq.~\ref{arc2} as 
\begin{equation}
{A_{ep}/A_0}  =
   Q_{W}^{p} + Q^2 B (Q^{2},\theta), 
\label{BTermEq}
\end{equation}
where $Q^{p}_{W}$ appears as the intercept, and the slope containing the hadronic structure is wrapped up in the B(Q$^2$, $\theta$) term. This latter term can be determined from existing PVES data at higher Q$^2$, and is quenched at small Q$^2$. In order to make use of Eq.~\ref{BTermEq}, it is assumed that the only significant energy dependent electroweak radiative correction $\Box^V_{\gamma {\rm Z}}({ E,Q^2})$ has first been subtracted from the asymmetry. Results of several recent calculations of this radiative correction are presented  in table~\ref{table1}. There is good agreement about the magnitude of the correction. The most recent and most precise calculation~\cite{WallyNew}  improved the precision through the use of parton distribution functions, and 
recent $\vec{\rm e}$d parity violation (PV) data from JLab~\cite{Zheng}. Their result corresponds to a 7.8\% $\pm$ 0.5\% correction at the kinematics of this experiment to the Standard Model (SM) value of $Q^{p}_{W}$ (0.0710(7))~\cite{PDG2012}.

\begin{table}
\centering
\caption{Recent calculations of $\Box^V_{\gamma {\rm Z}}({ E,Q^2})$ and its uncertainty at the kinematics of this measurement.}
\label{table1}      
\begin{tabular}{lll}
\hline
Reference & $\Box^V_{\gamma {\rm Z}}({ E,Q^2})$  & $\Delta \Box^V_{\gamma {\rm Z}}({ E,Q^2})$  \\\hline
Gorchtein, et al.~\protect{\cite{Gorchtein1}} & 0.0026 & 0.0026 \\
Sibirtsev, et al.~\cite{Wally4} & 0.0047 & $^{+0.0011}_{-0.0004}$ \\
Rislow, et al.~\cite{Rislow} & 0.0057 & 0.0009 \\
Gorchtein, et al.~\cite{Gorchtein2} & 0.0054 & 0.0020 \\
Hall, et al.~\cite{WallyNew} & 0.0056 & 0.00036 \\ \hline
\end{tabular}
\end{table}

\section{The Experiment}
\label{experiment}
\begin{figure}[h]
\centering
\sidecaption
\includegraphics[width=10cm,clip]{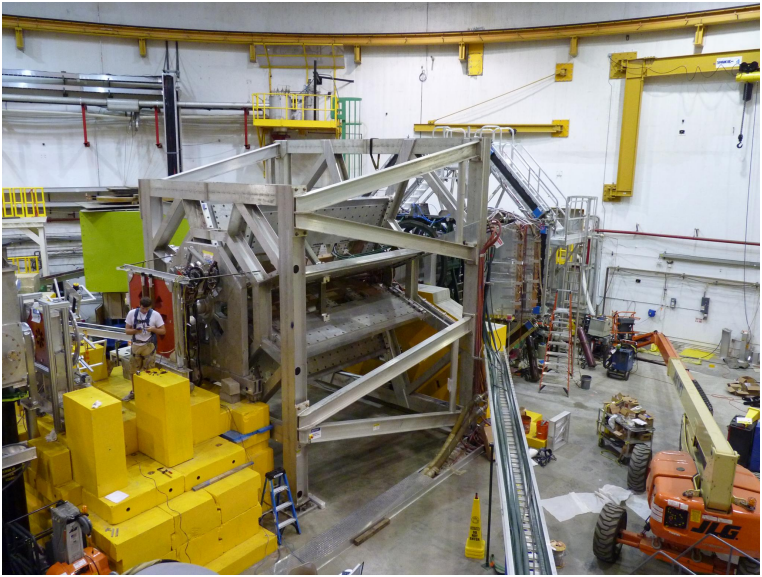}
\caption{Photograph of the experiment during installation. The beam travels from left to right through the target scattering chamber (partially visible at left), the collimation region, the toroidal magnet, and into the quartz detector bars arrayed octagonally around the beam axis just downstream of the magnet.}
\label{installationpic}      
\end{figure}

A dedicated apparatus was constructed for this experiment~\cite{proposal} at JLab. The main components were a 35 cm long LH$_2$ target, a triple lead collimator system to define the acceptance, and a toroidal magnet used to separate elastic events from inelastic events at a focus where eight quartz Cerenkov detectors were arrayed around the beam axis.  Retractable wire chambers~\cite{Leckey} were situated before and after the magnet to characterize the Q$^2$ of the experiment. The regions between the target and the magnet, as well as the detector region, were heavily shielded. The experiment is shown part way through its installation in Fig.~\ref{installationpic}, before the shielding was in place. 

The commissioning phase of the experiment reported here made use of a 1.155 GeV electron beam with longitudinal polarization of 89\% $\pm$ 1.8\%. The beam current was 145$-$180 $\mu$A. The mean scattering angle was 7.9$^\circ$ with an acceptance width of about $\pm$3$^\circ$. The azimuthal acceptance was nearly half of 2$\pi$. The experiment's Q$^2$ was determined via simulation to be 0.0250 $\pm$ 0.0006 GeV$^2$.

\subsection{Polarimetry}
\label{polarimetry}
After statistics, the next largest contribution to the uncertainty on the asymmetry is expected to come from the determination of the beam polarization. An existing M{\o}ller polarimeter~\cite{HallCMoller} routinely provides percent-level precision in JLab's Hall C. The polarimeter makes use of known analyzing powers provided by a fully polarized iron foil in a 3.5~T field. However, the measurement is invasive to the main experiment, and can only be performed at low beam currents. Therefore, a new Compton polarimeter was built for this experiment to complement the M{\o}ller polarimeter with a continuous, non-invasive and high current 1\%/hour device. A circularly polarized green laser in a low gain cavity provides the known analyzing power. The agreement between the two polarimeters is well within their uncertainties.

\subsection{Target}
\label{target}
The most challenging component of the experiment was the liquid hydrogen target~\cite{targetPAVI11}. It had to satisfy the mutually opposing requirements of simultaneously being the highest power target in the world (>2100 W of beam power at 180 $\mu$A) while also providing the smallest density fluctuations ever achieved (< 50 ppm). The LH$_2$ was circulated in a closed loop by means of a centrifugal pump which provided a head of 1.1 psi at the design capacity of 15 l/s (1.1 kg/s). The LH$_2$ flow was directed transversely across the beam axis in the 34.5 cm long  cell. All the scattered electrons in the experiment's acceptance passed perpendicularly through the larger diameter convex exit window  of the conical cell. The other elements of the loop were  a 3 kW resistive heater, and a 3 kW counterflow hybrid heat exchanger which simultaneously made use of helium coolant supplied at 1.2 MPa and 14~K as well as 0.3 MPa and 4 K in order to achieve the required 3 kW overall cooling power.  The target was held at 20.00 K and 0.22 MPa. 
The target cell was designed using computational fluid dynamics simulations in order to find the optimal geometry which minimized density fluctuations of the liquid along the beam axis of the conical cell, especially near the aluminum entrance and exit windows, which were 0.10 mm and 0.13 mm thick, respectively.  

\begin{figure}[h]
\centering
\sidecaption
\includegraphics[width=8cm,clip]{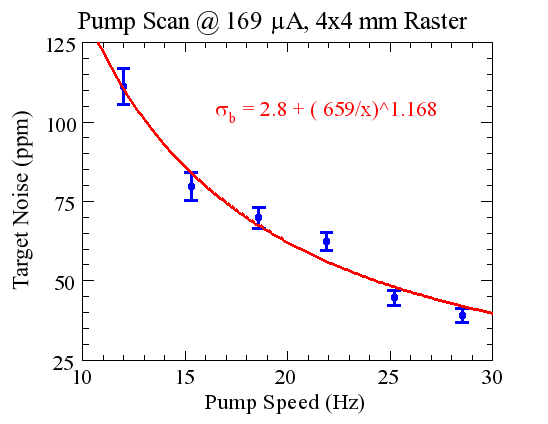}
\caption{Variation of target noise with the target pump speed extracted from the measured variation of the quartz detector asymmetry width $\Delta A_{\rm quartet}$, assuming the target noise contributed to $\Delta A_{\rm quartet}$ in quadrature with a constant term. The red curve is a fit to the measured data and indicates that the target noise falls off only slightly faster than the inverse of the pump speed.
}
\label{pumpscan}       
\end{figure}

The measured density fluctuations were only 37 $\pm$ 5 ppm with 169 $uA$ of beam dithered to a spot 4 x 4 mm$^2$ at the entrance to the target cell, and the target pump running at its nominal 28.5 Hz. This represented a very small part of the overall 236 ppm asymmetry width at this beam current. 
To help mitigate target noise, the beam polarization was reversed at 960 Hz for this experiment instead of the usual 30 Hz. The asymmetry width $\Delta A_{\rm quartet}$ was measured over helicity quartets ($\pm$$ \mp$$ \mp$$ \pm$). 
The statistical power of the experiment is proportional to $\Delta A_{\rm quartet}/\sqrt{N_{\rm quartets}}$. The measured contribution of the target noise to the asymmetry width in the experiment was determined in 3 independent ways, by varying either the beam current, the size of the beam at the target, or the speed of the pump which circulated the hydrogen across the beam axis. A plot showing the results of one of the pump speed variation studies is shown in Fig.~\ref{pumpscan}. 

\subsection{Detectors}
\label{detectors}
Eight synthetic quartz detectors~\cite{Peiqing} were  symmetrically arrayed around  the beam axis at a radius of 3.4 m,  12.2 m downstream of the target. The azimuthal symmetry of the detectors helped to reduce errors from helicity-correlated beam motion and transverse beam polarization. Each of the 8 detectors were formed from 1 m long bars glued end to end. The resulting 2 m long bars were 18 cm wide and 1.25 cm thick, and were fronted with 2 cm of lead pre-radiator. Low gain 12.7 cm  PMTs viewed the detectors via 18 cm long light guides at each end. The PMT anode current fed custom low noise I to V preamplifiers whose signals were digitized with  18 bit ADCs sampling at 500 kHz. During dedicated low current (0.1 - 200 nA) studies, different bases were used with the PMTs so that individual pulses could be counted along with the information from the drift chambers before and after the magnet. These low current studies were used to measure the Q$^2$ of the experiment, and to determine the detector response across the detector bars~\cite{Leckey}.

\section{Analysis}
\label{analysis}

The measured asymmetry was constructed from the charge normalized $\vec{\rm e}$p yields $Y^\pm$ according to 

\begin{equation}
A_{msr}  = \frac{Y^+ - Y^-}{Y^+ + Y^-} + A_T - A_{reg}
\label{amsr}
\end{equation}

\noindent where $A_T$ is the remnant transverse asymmetry explicitly measured with transversely polarized beam, and the regression correction $A_{reg}$ accounts for false asymmetries measured with natural and driven beam motion for x, y, x$^\prime$, y$^\prime$, and beam energy. The charge asymmetry was driven to zero with a feedback loop. Backgrounds were accounted for with explicit measurements of each of four background asymmetries A$_i$ and their dilutions f$_i$. The backgrounds arose from the aluminum target cell windows, the beamline, soft neutral background, and inelastic events. The largest background was from the target cell windows, where the measured dilution was 3.2\% and the measured asymmetry for this background was 1.76 ppm. The final asymmetry was obtained from
\begin{equation}
{\rm A_{\it ep}} = R_{tot}\frac{ A_{\rm{\it msr}} / P  - \sum\limits^4_{i=1} f_i A_i }{1 - \sum f_i } . \label{eqn:acor}
\end{equation}
Here $R_{tot}  = 0.98$ accounts for the combined effects of radiative corrections, the non-uniform light and Q$^2$ distribution across the detectors, and corrections for the uncertainty in the determination of Q$^2$. P represents the measured beam polarization of 0.890 $\pm$ 0.018. The total dilution $f_{tot} = \sum f_i = 3.6\%$. The final corrected asymmetry from the commissioning data reported here~\cite{Rakitha}, comprising only about 4\% of the data obtained in the experiment, is 
A$_{\rm{\it ep}} = - 279 \pm 35~\mbox{(statistics)} \pm 31~\mbox{(systematics)}$~ppb.

\section{Results}
\label{results}

The result from the commissioning data  reported here was combined with other PVES results~\cite{SAMPLE, SAMPLEbkwrd,   Happex1, Happex1p1, Happex1p2, Happex2He, Happex3, G0forward, G0backward,     PVA41, PVA42, PVA43}   on hydrogen, deuterium, and helium in a global fit following the prescription in~\cite{2007C1paper}. All PVES data up to 0.63 GeV$^2$ were used. Five free parameters were varied in the fit: the weak charges C$_{1u}$ and C$_{1d}$, the strange charge radius $\rho_s$ and magnetic moment $\mu_s$, and 
the isovector axial form factor $G^{Z \;(T=1)}_{A}$. The isoscalar $G^{Z \;(T=0)}_{A}$ was constrained by theory~\cite{Zhu}. All the data were corrected for the energy dependence of the 
 $\gamma$-Z box diagram calculated in Ref.~\cite{WallyNew}. The  small $Q^2$ dependence of  the $\gamma$-Z box diagram above $Q^2$=$ 0.025 \;  {\rm (GeV)} ^2$ was included using the prescription provided in Ref.~\cite{Gorchtein2} with EM form factors from Ref.~\cite{KellyFFs}.
To illustrate the fit, the $\theta$ dependence of the data was removed using Eq.~\ref{arc2}, and the asymmetries were divided by A$_0$ (defined in Eq.~\ref{arc3}). The resulting plot conforms to Eq.~\ref{BTermEq} and illustrates the quality of the global fit. 
The intercept of the fit at $Q^2=0$ is 
$Q^{p}_W ({\rm PVES})$=$0.064 \pm 0.012$.

 \begin{figure}[htb] 
\centering
\sidecaption
 \hspace*{-1.5 mm}
{\resizebox{3.4in}{2.5in}{\includegraphics{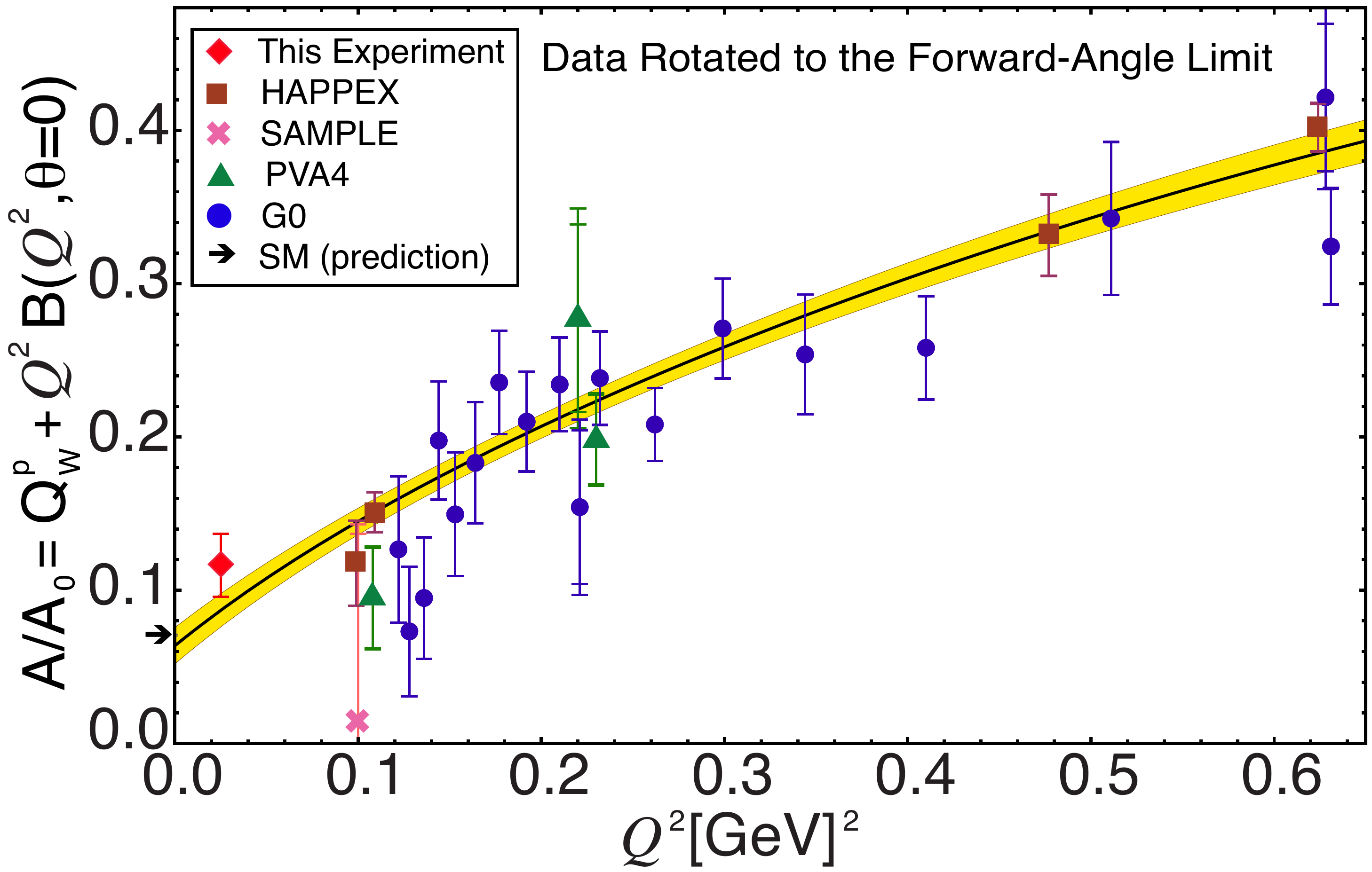}}}
 \caption{\label{BTerm}   
Global fit result (solid line) presented in the forward angle limit 
derived from this measurement as well as other PVES
experiments up to $Q^2 = 0.63$ $(GeV)^2$, including  proton, helium and deuterium data.
The additional uncertainty arising from the rotation is indicated by  outer error bars on each point, visible only for the more backward angle data. 
The yellow shaded region indicates the uncertainty in the fit. $Q^{p}_{W}$ is the intercept of the fit. 
The SM prediction~\cite{PDG2012} is also shown (arrow). }
 \end{figure}

As described in Sect.~\ref{intro}, the weak charge of the quarks can be extracted by combining this result with measurements of the weak charge on other targets. An especially precise measure of the weak charge of $^{133}$Cs has been reported~\cite{APV1} which serves this purpose. The most recent atomic corrections to this result are those of \cite{Flambaum2012}. Combining our result with the corrected APV result yields 
$C_{1u}$=$-0.1835 \pm 0.0054$ and $ C_{1d}$=$0.3355 \pm 0.0050$, with a correlation coefficient -0.980. 
Combining the $C_1$'s to extract the neutron's weak charge yields
$Q^{n}_W({\rm PVES\!+\!APV})$=$ -2(C_{1u} + 2C_{1d})$=$-0.975 \pm 0.010$. 
Both $Q^{p}_{W}$ and $Q^{n}_{W}$ are in agreement with the SM values~\cite{PDG2012} $Q^{p}_W({\rm SM})= 0.0710 \pm 0.0007$ and $Q^{n}_W({\rm SM})= -0.9890 \pm 0.0007$. 

The commissioning results reported here are derived from only about 4\% of the data that were collected for the full experiment. The full results should  be available in late 2014.

This work was supported by DOE contract No. DE-AC05-06OR23177, under which Jefferson Science Associates, LLC operates Thomas Jefferson National Accelerator Facility. 

\begin{figure}[bbbt]
\centering
{\resizebox{3.3in}{3.0in}{\includegraphics{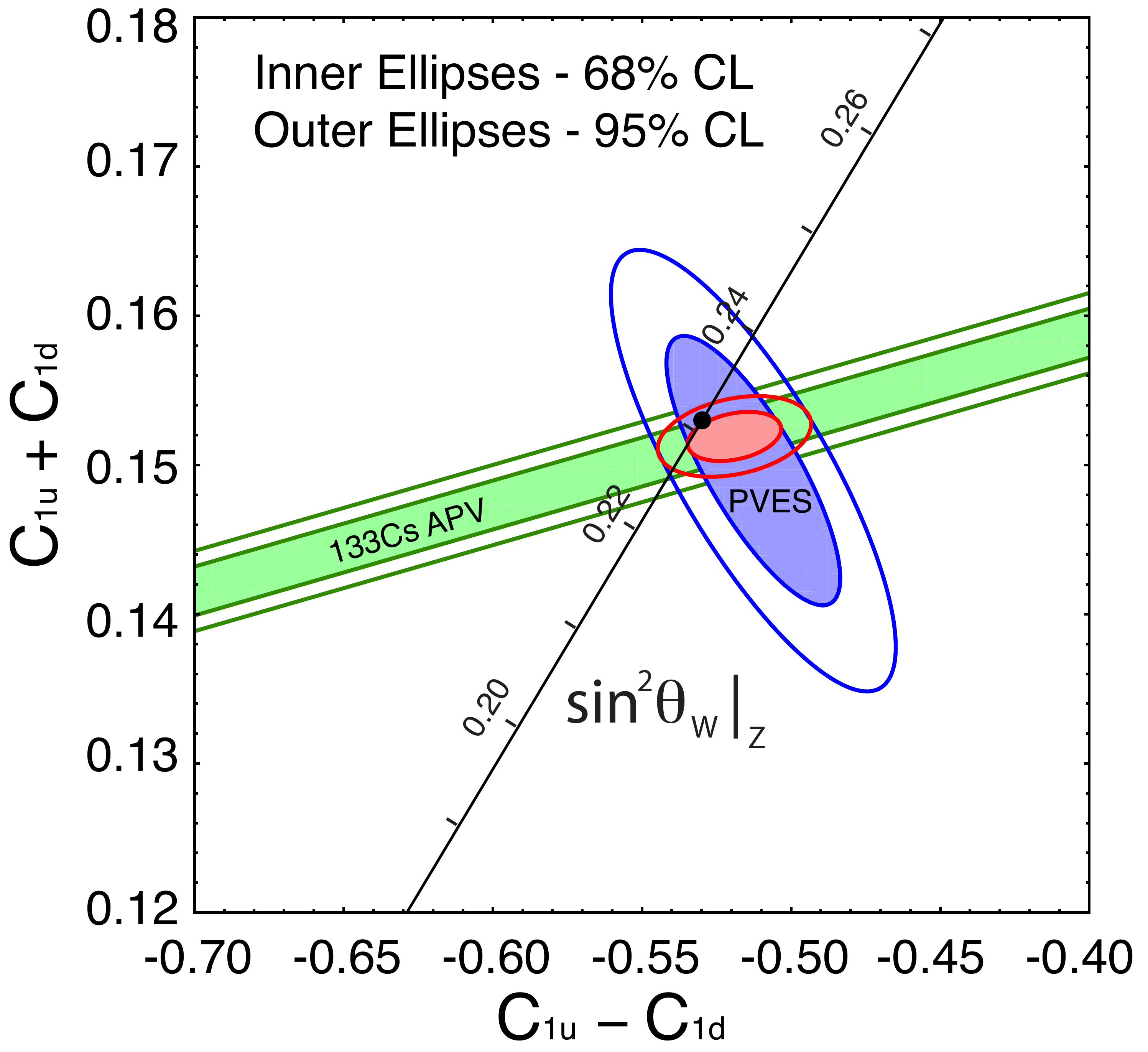}}}
 \caption{\label{C1plot}   Constraints on the neutral-weak quark coupling constants {$C_{1u}- C_{1d}$} (isovector) and 
{$C_{1u} + C_{1d}$} (isoscalar). 
The near horizontal (green) APV band constrains on the isoscalar combination from $^{133}$Cs data.
The vertical (blue) ellipse 
represents the global fit of the existing $Q^2 < 0.63$ $({\rm GeV})^{2}$ PVES data including the new result reported here at $Q^{2}$=0.025 $({\rm GeV})^{2}$.
The small (red) ellipse near the center of the figure shows the result obtained by combining the APV and PVES information. 
The SM prediction~\cite{PDG2012} as a function of $\sin^{2}\theta_W $ in the $\overline{MS}$
scheme is plotted (diagonal black line) with the SM best fit value indicated by the (black) point at $\sin^{2}\theta_W $=0.23116. 
}
 \end{figure}




%
%
%

\end{document}